\documentclass[preprint,showpacs,preprintnumbers,amsmath,amssymb]{revtex4}

\def\undersim#1{\setbox9\hbox{${#1}$}{#1}\kern-\wd9\lower
    2.5pt \hbox{\lower\dp9\hbox to \wd9{\hss $_\sim$\hss}}}
\usepackage{amssymb}
\usepackage{amsmath}
\usepackage{graphicx}
\usepackage{color}

\def\undersim#1{\setbox9\hbox{${#1}$}{#1}\kern-\wd9\lower
    2.5pt \hbox{\lower\dp9\hbox to \wd9{\hss $_\sim$\hss}}}

\def\mk{{\mathbf p}}

\begin{document}

\title{Squeezed spectra and elliptic flow of bosons and anti-bosons with in-medium mass splitting}

\author{Yong Zhang$^{1}$}
\author{Shi-Yao Wang$^{2}$}
\author{Peng Ru$^{3,}$}
\author{Wei-Hua Wu$^{1}$}
\affiliation{\footnotesize $^1$School of Mathematics and Physics, Jiangsu University of Technology, Changzhou, Jiangsu 213001, China\\
$^2$School of Physics, Dalian University of Technology, Dalian, Liaoning 116024, China\\
$^3$School of Materials and New Energy, South China Normal University, Shanwei 516699, China}
%\date{\today}

\begin{abstract}
We study the impact of the in-medium mass splitting between bosons and anti-bosons on their spectra and elliptic flow.
The in-medium mass splitting may cause a separation in the transverse momentum spectra, as well as a division in the elliptic flow between bosons and anti-bosons.
The magnitude of this effect becomes greater as the in-medium mass splitting increases.
With the increasing rapidity, the splitting effect of the spectra
increases and the splitting effect of the elliptic flow decreases.
These phenomena may provide a way to differentiate whether the influences on boson and anti-boson in the medium are consistent.


\end{abstract}

\maketitle

\section{Introduction}
In high-energy heavy-ion collisions, people study the properties of the system and the matter created in collisions by analyzing
the information of detected particles. The transverse momentum spectra and elliptic flow $v_2$ of detected particles have been extensively studied
in such collisions \cite{STAR_PRL04_TMS,PHENIX_PRC04_TMS,ALICE_PRL12_TMS,ALICE_PRC13_TMS,STAR-v2-05,PHENIX-v2-09,ALICE-v2-15,Heinz:2009xj,Ollitrault:1992bk,
Schenke:2011qd,Snellings:2014kwa}.

Particles undergo interactions with the surrounding medium before being detected and may cause the
in-medium mass of the particles to be unequal to their mass in vacuum. The in-medium mass modification may lead
to a squeezing effect, and directly lead to a new experimental observation between boson and anti-boson, called
squeezed back-to-back correlation \cite{AsaCso96,AsaCsoGyu99,Padula06,Zhang15a}.
The transverse momentum spectra and elliptic flow $v_2$ may also affected by this effect \cite{AsaCsoGyu99,Padula06,Zhang20,Zhang21}.
Generally, the medium effects on charged particles and their antiparticles are inconsistent \cite{XPZ-19}. Thus,
the medium mass of the charged particles and their antiparticles may be different.
The medium effects on transverse momentum spectra of boson and anti-boson was investigated for spherical expanding Gaussian source.
In-medium mass splitting may cause a separation in the transverse momentum spectra of boson and anti-boson \cite{Zhang21}.
This phenomenon may provide evidence to support that the medium effects on boson and anti-boson are not the same.
The in-medium mass splitting may also lead to different effects on elliptic flow $v_2$ of boson and anti-boson.
In this paper, a more realistic model will be applied to study the different medium effects on
the transverse momentum spectra and elliptic flow $v_2$ of boson and anti-boson. In this paper, the
transverse evolution of the system is described by the ideal relativistic hydrodynamics in $2+1$ dimensions,
and the Bjorken boost-invariant hypothesis \cite{Bjo83} is used to describe the longitudinal evolution of the system.
This model is suitable from the RHIC top energy to the LHC energy \cite{{Ris98,KolHei03,Bay83,Gyu97,Ris9596,BLM96,
BLM04,Kol00,KolRap03,She10,ZhaEfa2,HuichaoS}}.

The different medium effects on the transverse momentum spectra and elliptic flow $v_2$ of $D$ meson and anti-$D$ meson
is shown in this paper. Since $D$ meson is produced before quark-gluon plasma (QGP) and it is one important probe for
studying the properties of QGP \cite{alice-D,FML-prc}. On the other hand, the measurement of $D$ meson has aroused great
interest \cite{STAR-PRL14,STAR-NPA16,ALICE-PRL13,ALICE-PRC14,ALICE-PRL18,ALICE-EPJC19,ALICE-EPJC20,ALICE-PLB21,ALICE-JHEP21}.

The remaining parts of this paper are planned as follows. The expressions for calculating
the momentum distributions of boson and anti-boson are presented in Sec. II. Then, the results of
different medium effects on the transverse momentum spectra and elliptic flow $v_2$
of $D$ meson and anti-$D$ meson are shown in Sec. III.
Finally, the summary and discussion are given in Sec. IV.

\section{Formulas}
The boson and anti-boson are affected by the surrounding medium, and they are treated as quasi-particles.
In this paper, the creation (annihilation) operator of the boson and anti-boson in medium are represented
by $a'^\dagger_\mk\, (a'_\mk)$ and $b'^\dag_\mk\,(b'_\mk)$ respectively, and $\mk$ represents
the momentum of the boson and anti-boson. The creation (annihilation) operator of the boson and anti-boson
in vacuum are represented by $a^\dagger_\mk\, (a_\mk)$ and $b^\dag_\mk\,(b_\mk)$ respectively.
Bogoliubov transformation was used to relate the creation (annihilation) operators of the free particles and
the quasi-particles \cite{XPZ-19},
\begin{equation}
a_\mk =\cosh f_\mk a'_\mk +\sinh f_-\mk b'^\dag_{-\mk},
\end{equation}
\begin{equation}
b_\mk =\cosh f_\mk b'_\mk +\sinh f_-\mk a'^\dag_{-\mk},
\end{equation}
where
\begin{equation}
f_\mk =\frac{1}{2} \ln\left({\omega_\mk}/{\Omega_\mk}\right),
\end{equation}
\begin{equation}
\omega_\mk =\sqrt{\mk^2 +m^2},~~~~\Omega_\mk =\sqrt{\mk^2 +(m+ \Delta m)^2 +\delta^2}.
\end{equation}
In Eq.\,(4), $\omega_\mk$ and $m$ represent the energy and the mass of free bosons with momentum $\mk$ respectively.
$\delta$ represents half of the energy difference between boson and anti-boson in the medium \cite{XPZ-19}. When the energy difference
is zero ($\delta\,=\,0$), the boson and anti-boson have the same in-medium mass modification $\Delta m (\mk)$, and it is taken as:
\begin{equation}
\Delta m (\mk) = \Delta m_0 \exp[-\mk^2/{\Lambda}_s^2],
\end{equation}
where ${\Lambda}_s$ is a parameter describing the momentum dependence of the mass modification, and $\Delta m_0$ is the
mass modification of the boson and anti-boson in the medium for $\mk = 0$.
For nonzero in-medium energy splitting ($\delta\,>\,0$), the masses of boson and anti-boson in the medium become \cite{XPZ-19}
\begin{equation}
m'_{a}= (\Omega_{\mk} + \delta)\big|_{\mk=0}=\sqrt{(m+ \Delta m_0)^2+\delta^2}
+ \delta,
\end{equation}
\begin{equation}
m'_{b}= (\Omega_{\mk} - \delta)\big|_{\mk=0}=\sqrt{(m+ \Delta m_0)^2+\delta^2}
- \delta.
\end{equation}
For a pair of quasi-particles in the medium, the sign "a" represents the one with a larger mass.
For fixed momentum, the mass splitting in the medium between the boson and anti-boson is twice as much as $\delta$.

The single-particle spectral functions of boson and anti-boson from hydrodynamic sources are \cite{AsaCsoGyu99,Padula06,Zhang15a,Zhang21,Cooper}
\begin{eqnarray}\label{SP}
 N_a(\mk)\!&&=\!\int \frac{g_i}{(2\pi)^3}d^4\sigma_{\mu}(r)p^\mu\, \! \Bigl\{|c'_{\mk'}|^2\,
n'_{a,\mk'}+\,|s'_{-\mk'}|^2\,[\,n'_{b,-\mk'}+1]\Bigr\},
\end{eqnarray}
\begin{eqnarray}\label{SP1}
 N_b(\mk)\!&&=\!\int \frac{g_i}{(2\pi)^3}d^4\sigma_{\mu}(r)p^\mu\, \! \Bigl\{|c'_{\mk'}|^2\,
n'_{b,\mk'}+\,|s'_{-\mk'}|^2\,[\,n'_{a,-\mk'}+1]\Bigr\},
\end{eqnarray}
\begin{equation}\label{csk}
c'_{\mk'}=\cosh[\,f'_{\mk'}\,], \,\,\,s'_{\mk'}=\sinh[\,f'_{\mk'}\,],
\end{equation}
\begin{eqnarray}
f'_{\mk'}=\frac{1}{2} \log \left[\omega'_{\mk'}/\Omega'_{\mk'}\right]
=\frac{1}{2}\log\left[p^{\mu}u_{\mu}(r)/p^{*\nu}u_{\nu}(r)
\right],
\end{eqnarray}
\begin{eqnarray}
&&\hspace*{-7mm}\Omega'_{\mk'}(r)=\sqrt{\mk'^2(r)+(m+ \Delta m)^2 +\delta^2}\nonumber\\
&&\hspace*{3.9mm}=\sqrt{[p^{\mu} u_{\mu}(r)]^2-m^2+(m+ \Delta m)^2 +\delta^2}\nonumber\\
&&\hspace*{3.8mm}=p^{*\mu} u_{\mu}(r),
\end{eqnarray}
\begin{equation}\label{BZ}
n'_{a,\mk'}=\frac{1}{\exp[(\Omega_{\mk'}(r)+\delta)/T(r)]-1},
\end{equation}
\begin{equation}\label{BZ1}
n'_{b,\mk'}=\frac{1}{\exp[(\Omega_{\mk'}(r)-\delta)/T(r)]-1}.
\end{equation}

In the simulations, the equation of state of s95p-PCE \cite{She10} is chosen.
The initial condition for the hydrodynamic simulation is set as Gaussian distribution
in the plane at $z = 0$ and $\tau_0 = 0.6$ fm/$c$ after nucleus-nucleus collisions:
\begin{equation}
\epsilon(x,y) = \epsilon_0\exp[-x^2/(2R_x^2)-y^2/(2R_y^2)],
\end{equation}
where $\epsilon(x,y)$ is the energy density at $(x,y)$ and $z = 0$, and $\epsilon_0$ is the energy density at the
center of the source. $R_x$ and $R_y$ represent the radii in the $x$ and $y$ directions, respectively \cite{Zhang15a}.

\section{Results}

\begin{figure}[htbp]
\includegraphics[scale=0.9]{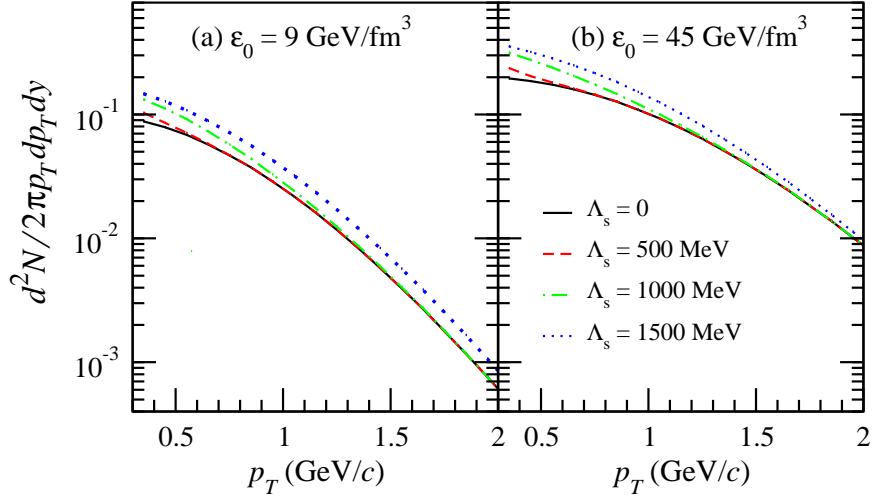}
\caption{(Color online) The transverse momentum spectra of $D$ meson with different ${\Lambda}_s$ for $\delta\,=\,0$. }
\label{Dsp}
\end{figure}

First, we consider the case that the in-medium mass of the boson and anti-boson are the same $(\delta~=~0)$.
In Fig.~\ref{Dsp}, the transverse momentum spectra of $D$ meson with various ${\Lambda}_s$ for $\delta\,=\,0$ are presented.
The temperature at which $D$ meson freezes out is considered to be 0.15 GeV \cite{XPZ-19} and the in-medium mass-shift
parameter $\Delta m_0$ is taken as $-\,5$ MeV \cite{XPZ-19,FMFK-PRC06,Yang-CPL18}. 
The initial radii, $R_x$ and $R_y$ are assumed to be 5 fm, while $\epsilon_0$ is considered as 9 and 45 GeV/fm$^3$ \cite{Zhang20}.
There is no medium effect for ${\Lambda}_s =0$. An increase in the yield of $D$ meson is observed due to the
modification of its in-medium mass. This effect increases with the increasing ${\Lambda}_s$.
\begin{figure}[htbp]
\includegraphics[scale=0.9]{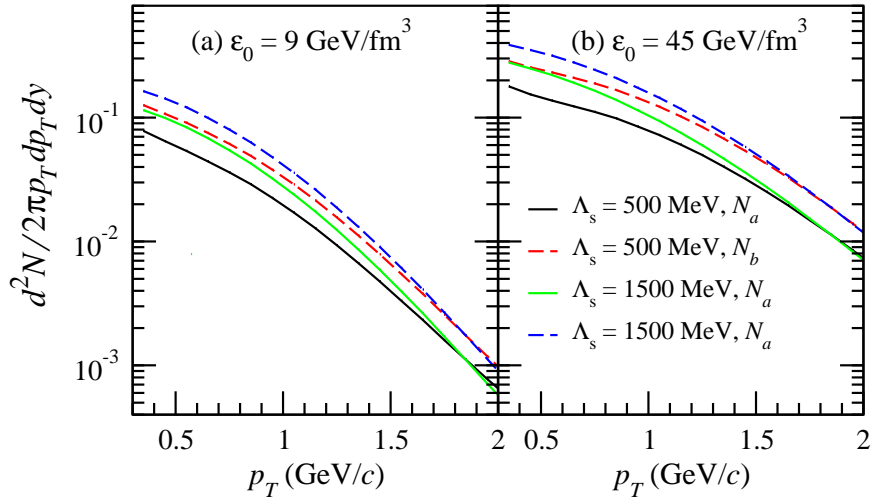}
\caption{(Color online) The transverse momentum spectra of $D$ meson and anti-$D$ meson for $\delta\,=\,40$ MeV.
Here $N_a$ represents the spectrum of the particle with a larger in-medium mass.}
\label{Dsp1}
\end{figure}

The transverse momentum spectra of $D$ meson and anti-$D$ meson are shown in Fig.~\ref{Dsp1} for $\delta\,=\,40$ MeV.
There is a separation in the transverse momentum spectra between $D$ meson and anti-$D$ meson for nonzero $\delta$.
\begin{figure}[htbp]
\includegraphics[scale=0.9]{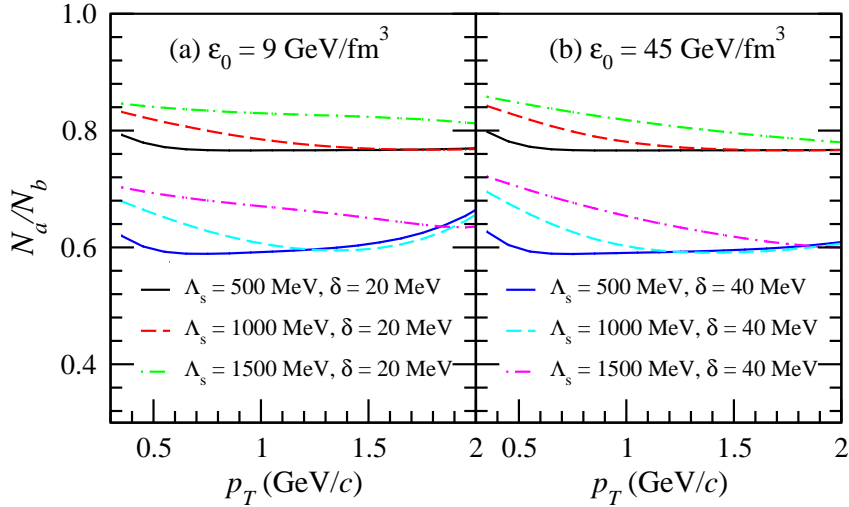}
\caption{(Color online) The ratio $N_a/N_b$ of $D$ meson for $\delta\,=\,20$ and $40$ MeV.}
\label{spratio}
\end{figure}
To quantitatively exhibit the impact of nonzero in-medium mass splitting on the transverse momentum spectra of $D$ meson and anti-$D$ meson,
we show the ratio $N_a/N_b$ of $D$ meson for $\delta\,=\,20$ and $40$ MeV in Fig. \ref{spratio}. For fixed ${\Lambda}_s$, the ratio $N_a/N_b$ declines as the
in-medium mass splitting increases. The ratio $N_a/N_b$ declines as the transverse momentum $p_T$ increases within the small transverse momentum region.

\begin{figure}[htbp]
\vspace*{0mm}
\includegraphics[scale=0.9]{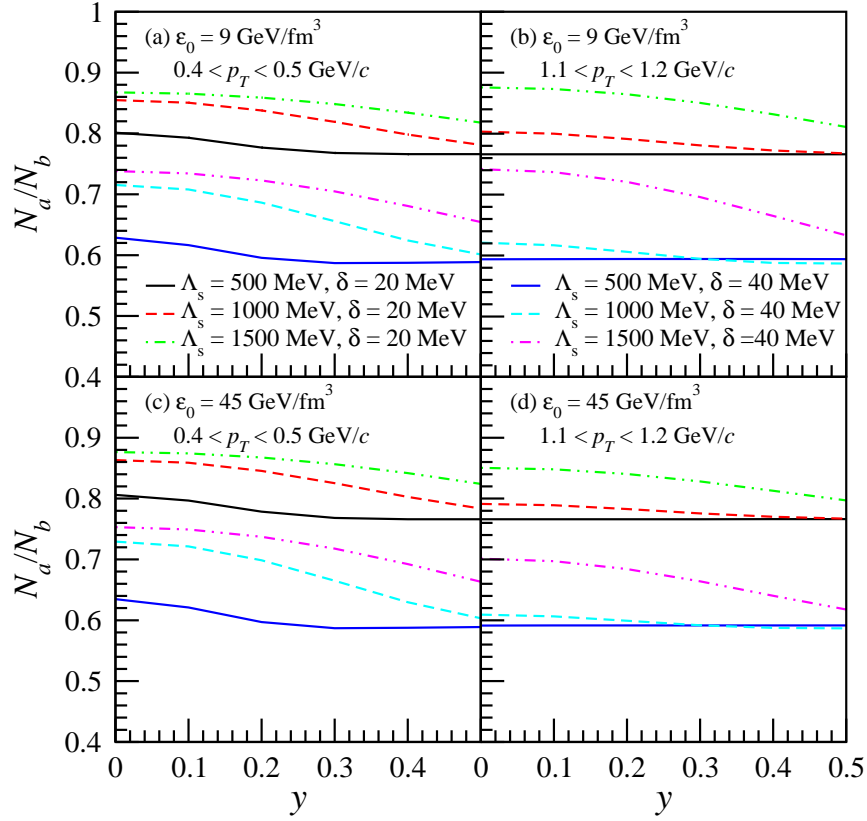}
\caption{(Color online) Dependence of the ratio $N_a/N_b$ of $D$ meson on the
rapidity of the particle.}
\label{spkd}
\end{figure}

In Fig. \ref{spkd}, we show the dependence of the ratio $N_a/N_b$ of $D$ meson on the
rapidity of the particle. The ratio $N_a/N_b$ declines with the increasing rapidity within the range of 0.4 GeV/$c$ $\leq$ $p_T$ $\leq$ 0.5 GeV/$c$.
For ${\Lambda}_s = 500$ MeV, the ratio $N_a/N_b$ does not depend on the rapidity within the range of 1.1 GeV/$c$ $\leq$ $p_T$ $\leq$ 1.2 GeV/$c$.
The ratio $N_a/N_b$ decreases with the increasing rapidity within the range of 1.1 GeV/$c$ $\leq$ $p_T$ $\leq$ 1.2 GeV/$c$ for ${\Lambda}_s = 1000$ MeV and $1500$ MeV.
According to the aforementioned findings, a separation in transverse momentum spectra between $D$ meson and anti-$D$ meson may arise due to the in-medium mass splitting. The amount of separation escalates with higher in-medium mass splitting as well as particle rapidity.

\begin{figure}[htbp]
\includegraphics[scale=0.9]{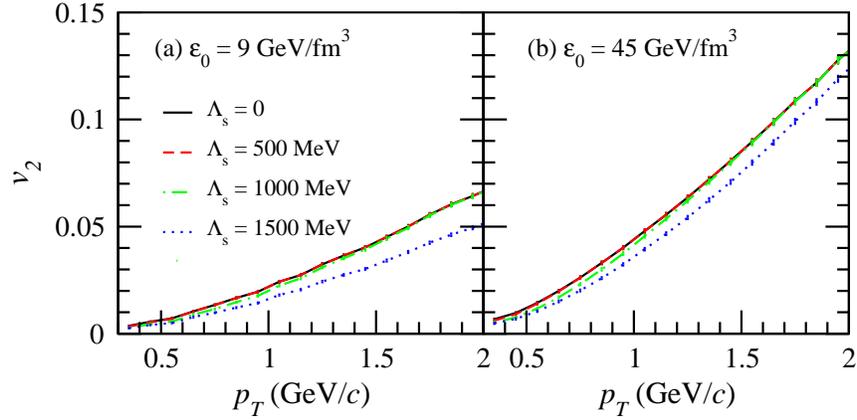}
\caption{(Color online) The elliptic flow $v_2$ of $D$ meson with various ${\Lambda}_s$ for $\delta\,=\,0$. }
\label{Dv2}
\end{figure}

\begin{figure}[htbp]
\hspace*{0mm}
\includegraphics[scale=0.9]{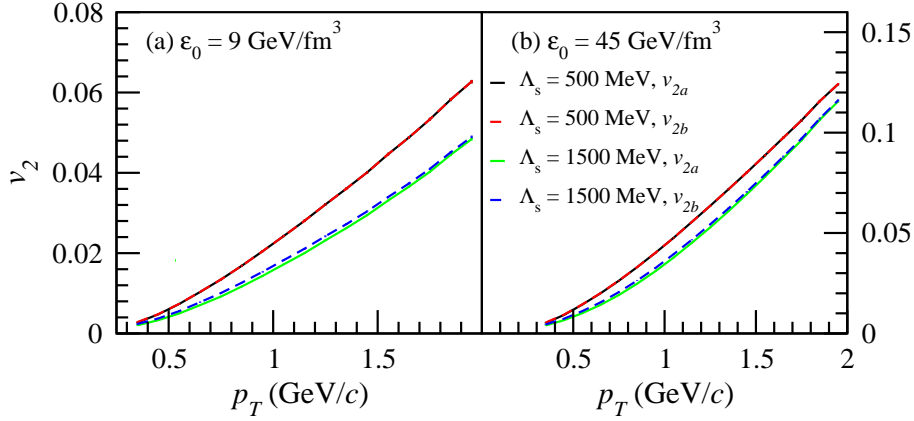}
\caption{(Color online) The elliptic flow $v_2$ of $D$ meson and anti-$D$ meson for $\delta\,=\,40$ MeV. Here $v_{2a}$ represents the elliptic flow of the particle with a larger in-medium mass. }
\label{Dvs2}
\end{figure}
\begin{figure}[htbp]
\hspace*{3mm}
\includegraphics[scale=0.9]{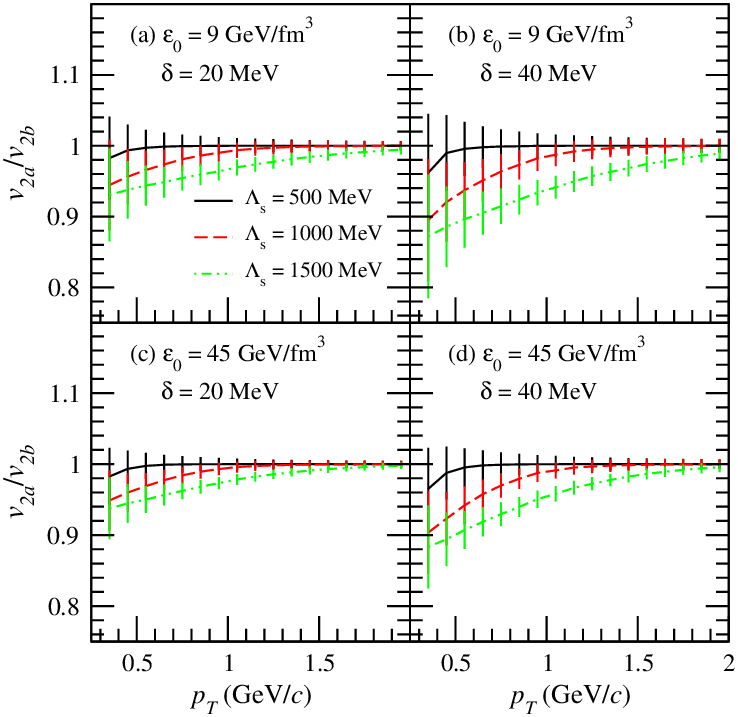}
\caption{(Color online)The ratio $v_{2a}/v_{2b}$ of $D$ meson for $\delta\,=\,20$ and $40$ MeV.}
\label{Dvratio}
\end{figure}

One of the greatest extensively researched observables in non-central nucleon-nucleon collisions is elliptic flow.
To study the impact of the mass splitting in the medium on
the elliptic flow $v_2$, the initial source radii $R_x$ and $R_y$ are taken 4 fm and 5 fm, respectively. The elliptic flow
$v_2$ of $D$ meson with various ${\Lambda}_s$ for $\delta\,=\,0$ are shown in Fig. \ref{Dv2}. The in-medium mass-shift suppresses the
elliptic flow and this effect becomes more pronounced as ${\Lambda}_s$ increases. The elliptic flow $v_2$ of $D$ meson and anti-$D$ meson for $\delta\,=\,40$ MeV
are shown in Fig. \ref{Dvs2}. Here $v_{2a}$ represents the elliptic flow of the particle with greater in-medium mass.
Due to the in-medium mass splitting, there is only a slight variation observed between the elliptic flow of $D$ meson and anti-$D$ meson.
In Fig. \ref{Dvratio}, we show the
ratio $v_{2a}/v_{2b}$ of $D$ meson for $\delta\,=\,20$ and $40$ MeV. The ratio $v_{2a}/v_{2b}$ decreases with the increasing in-medium mass splitting $\delta$
and declines as the ${\Lambda}_s$ increases. For fixed $\delta$ and ${\Lambda}_s$, the variation between $v_{2a}$ and $v_{2b}$ reduces with the
increasing transverse momentum.

\begin{figure}[htbp]
\vspace*{0mm}
\includegraphics[scale=0.9]{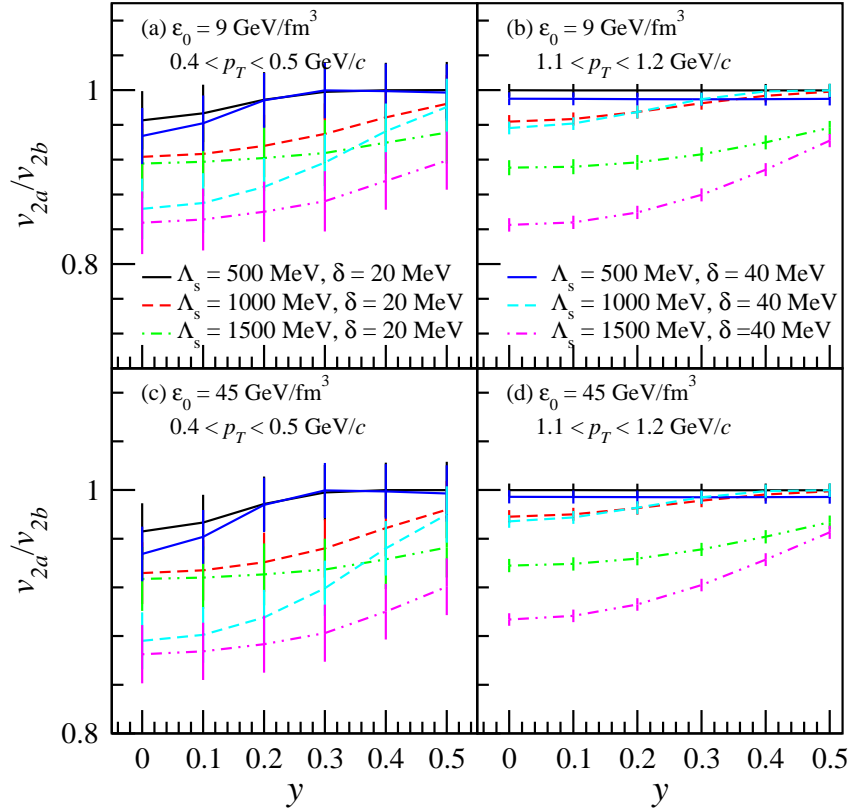}
\caption{(Color online) Dependence of the ratio $v_{2a}/v_{2b}$ of $D$ meson on the
rapidity of the particle.}
\label{vkd}
\end{figure}

In Fig. \ref{vkd}, the dependence of the ratio $v_{2a}/v_{2b}$ of $D$ meson on the
rapidity of the particle is presented. The ratio $v_{2a}/v_{2b}$ increases with the increasing rapidity within the range of 0.4 GeV/$c$ $\leq$ $p_T$ $\leq$ 0.5 GeV/$c$.
For ${\Lambda}_s = 500$ MeV, the ratio $v_{2a}/v_{2b}$ does not depend on the rapidity within the range of 1.1 GeV/$c$ $\leq$ $p_T$ $\leq$ 1.2 GeV/$c$.
The ratio $v_{2a}/v_{2b}$ increases with the increasing rapidity within the range of 1.1 GeV/$c$ $\leq$ $p_T$ $\leq$ 1.2 GeV/$c$ for ${\Lambda}_s = 1000$ MeV and $1500$ MeV.
The results in Fig. \ref{Dv2}\,--\,\ref{vkd} suggest that the in-medium mass splitting can cause a division of the elliptic flow between $D$ meson and anti-$D$ meson. As the in-medium mass splitting increases, this division becomes more noticeable, yet it diminishes as the rapidity increases.

\section{Summary and discussion}
The mass modification of bosons in the particle-emitting source produced in high-energy heavy-ion collisions can lead to a squeezing effect due to interactions with the medium.
This effect may have an impact on the transverse momentum spectra and elliptic flow of particles.
Charged particles and their antiparticles exhibit distinct interactions within a medium.
This paper explores how in-medium mass disparity affects the spectra and elliptic flow of bosons and anti-bosons.
Our results indicate that the in-medium mass splitting may cause a separation in the transverse momentum spectra,
as well as a division in the elliptic flow between bosons and anti-bosons.
Moreover, the magnitude of this effect becomes greater as the in-medium mass splitting increases.
At larger rapidity, the difference is larger in the spectra and smaller in elliptic flow $v_2$. These phenomena may provide new means to
differentiate whether the influences on boson and anti-boson in the medium are consistent.

In the calculations, we introduce two parameters to describe the in-medium mass modification.
The momentum-dependent $\Delta m$ signifies the equivalent portion of the in-medium mass-shift for both the boson and anti-boson.
The energy splitting between the boson and anti-boson in the medium is represented by $\delta$, and it is considered independent of momentum.
If the $\delta$ decreases as momentum increases, the splitting effect may be important for boson and anti-boson in small momentum regions.

With the increasing collision energy, the bosons are subject to a more intense medium effect. The
squeezed back-to-back correlation between boson and anti-boson is sensitive to particle emission time distribution \cite{Padula06,Zhang15a,zhang_epjc},
and it may be suppressed to no observed signal by the wide particle emission time distribution. Thus, the
splitting effect of the spectra or elliptic flow of bosons and anti-bosons may provide another means for studying the
interactions between the bosons and the medium for the sources with a wide temporal distribution.

\begin{acknowledgments}
This research was supported by the National Natural Science Foundation
of China under Grant No. 11905085 and Changzhou Science and Technology Bureau under Grant No. CJ20210150.
\end{acknowledgments}

\end{document}